\documentstyle[12pt]{article}
\begin{document}
\begin{center}{\large{\bf Mapping Among Manifolds II}}
\end{center}
\vspace*{1.5cm}
\begin{center}
A. C. V. V. de Siqueira
$^{*}$ \\
Departamento de Educa\c{c}\~ao\\
Universidade Federal Rural de Pernambuco \\
52.171-900, Recife, PE, Brazil\\
\end{center}
\vspace*{1.5cm}
\begin{center}{\bf Abstract}

In a previous paper we built a modified Hamiltonian formalism to
make possible explicit maps among manifolds. In this paper the
modified formalism was generalized. As an application, we have
built maps among spaces associated to spinors, as well as maps
among Kaehler spaces.
\end{center}

 \vspace{3cm}

${}^*$ E-mail: acvvs@ded.ufrpe.br
\newline

\newpage

\section{Introduction}
$         $

 The Jacobi fields are very important to the Riemannian Geometry
[1] and to the singularity theorems [2],[3]. These fields were
used to study a free falling particle motion in a Schwarzschild
spacetime [4], and a charged particle motion in Kaluza-Klein
manifolds [5]. In a previous paper we have modified the
Hamiltonian formalism to build  maps among manifolds [6]. In this
paper we present a second map building method among manifolds. As
an application, we have built maps among complex spaces, and also
among spaces associated to spinors.

 This paper is organized as follows.  In Sec. $2$ we give a brief
overview of the modified Hamiltonian formalism which we call
\emph{the first modified Hamiltonian formalism}. In Sec. $3$ we
build \emph{the second modified Hamiltonian formalism}. In Sec.
$4$ we apply this formalism to spaces associated to spinors. In
Sec. $5$ we apply it to complex spaces. In Sec. $6$ we summarize
the main results of this work.

\renewcommand{\theequation}{\thesection.\arabic{equation}}
\section{\bf The First Modified Hamiltonian Formalism}
\setcounter{equation}{0} $         $

 It is well-known that in the Hamiltonian formalism the
Hamilton equations and the Poisson brackets are conserved only by
a canonical or sympletic transformation. In [4] we changed the
non-relativistic time-dependent harmonic oscillator [7],[8] to a
general relativistic approach. In the first modified Hamiltonian
formalism only Hamilton equations will be kept, in the sense that
they will be transformed into other Hamilton equations by a
non-canonical or non-sympletic transformation, and the Poisson
brackets will not be invariant.

We now give a brief overview of the first modified Hamiltonian
formalism [6]. Consider a time-dependent Hamiltonian $H({\tau})$
where ${\tau}$ is an affine parameter, in this case, the
proper-time of the particle. Let us define 2n variables that will
be called ${\xi}^j$ with index j running from 1 to 2n so that we
have ${\xi}^j$ $\in$
$({\xi}^1,\ldots,{\xi}^n,{\xi}^{n+1},\ldots,{\xi}^{2n})$ =$(
{q}^1,\ldots,{q}^n,{p}^1,\ldots,{p}^n)$ where ${q}^j$ and ${p}^j$
can be or not coordinates and momenta, respectively. We now define
the Hamiltonian by
\begin{equation}
 H({\tau})=\frac{1}{2}H_{ij}{\xi}^i{\xi}^j,
\end{equation}
where $H_{ij}$ is a  symmetric matrix. We consider that the
Hamiltonian obeys the Hamilton equation
\begin{equation}
\frac{d{\xi}^i}{d\tau }={J}^{ik}\frac{\partial{H}}{\partial{\xi}^k
} .
\end{equation}
The equation (2.2) introduces the sympletic  J, given by
\begin{equation}
\left(%
\begin{array}{cc}
  O & I \\
  -I & O \\
\end{array}%
\right)
\end{equation}
where O and I are the $n \textbf{x}n$ zero and identity matrices,
respectively. We now make a linear transformation from ${\xi}^j$
to ${\eta}^j$ given by
\begin{equation}
  {\eta}^j={{T}^j}_k{\xi}^k,
\end{equation}
where ${{T}^j}_k$ is  a non-sympletic matrix, and the new
Hamiltonian is given by
\begin{equation}
 \bar{H}=\frac{1}{2}C_{ij}{\eta}^i{\eta}^j,
\end{equation}
where $C_{ij}$ is a  symmetric matrix. The matrices H, C, and T
obey the following system
\begin{equation}
 \frac{d{{T}^i}_j}{d\tau}+\frac{d{t}}{d\tau}{{T}^i}_k{J}^{kl}X_{lj}=J^{im}Y_{ml}{{T}^j}_k,
\end{equation}
where $2X_{lj}=\frac{\partial{H_{ij}}}{\partial{\xi}^l
}\xi^{i}+2H_{lj}$ and
$2Y_{ml}=\frac{\partial{C_{il}}}{\partial{\eta}^m
}\eta^{i}+2C_{ml},$ t and $\tau$ are the proper-times of the
particle in two different manifolds. We note that (2.6) is a first
order linear differential equation system in ${{T}^i}_k ,$ and it
is the response for what we looked for because the non-linearity
in the Hamilton equations were  transferred to their coefficients
[6]. Consider $\frac{d{t}}{d\tau}X_{lj}=Z_{lj}$ and write (2.6) in
the matrix form
\begin{equation}
 \frac{d{T}}{d\tau}+TJZ=JYT,
\end{equation}
where T, Z and Y are  $2n \textbf{x}2n$ matrices as
\begin{equation}
\left(%
\begin{array}{cc}
  T_{1} & T_{2} \\
  T_{3} & T_{4} \\
\end{array}%
\right)
\end{equation}
with similar expressions for Z and Y. Let us  write (2.7) as
follows
\begin{equation}
 \dot{T_1}=Y_{3}T_{1}+Y_{4}T_{3}+T_{2}Z_{1}-T_{1}Z_{3},
\end{equation}
\begin{equation}
 \dot{T_2}=Y_{3}T_{2}+Y_{4}T_{4}+T_{2}Z_{2}-T_{1}Z_{4},
\end{equation}
\begin{equation}
 \dot{T_3}=-Y_{1}T_{1}-Y_{2}T_{3}+T_{4}Z_{1}-T_{3}Z_{3},
\end{equation}
\begin{equation}
 \dot{T_4}=-Y_{1}T_{2}-Y_{2}T_{4}+T_{4}Z_{2}-T_{3}Z_{4}.
\end{equation}
Now consider
\begin{equation}
 \dot{S_1}=Y_{3}S_{1}+Y_{4}S_{3},
\end{equation}
\begin{equation}
 \dot{S_2}=Y_{3}S_{2}+Y_{4}S_{4},
\end{equation}
\begin{equation}
 \dot{S_3}=-Y_{1}S_{1}-Y_{2}S_{3},
\end{equation}
\begin{equation}
 \dot{S_4}=-Y_{1}S_{2}-Y_{2}S_{4},
\end{equation}
and
\begin{equation}
 \dot{R_1}=R_{2}Z_{1}-R_{1}Z_{3},
\end{equation}
\begin{equation}
 \dot{R_2}=R_{2}Z_{2}-R_{1}Z_{4},
\end{equation}
\begin{equation}
 \dot{R_3}=R_{4}Z_{1}-R_{3}Z_{3},
\end{equation}
\begin{equation}
 \dot{R_4}=R_{4}Z_{2}-R_{3}Z_{4}.
\end{equation}
From the theory of  first order differential equation systems [9],
  it is well-known that each system in (2.13)-(2.20) has a solution in the
  region where $Z_{lj}$ and $Y_{ml}$ are continuous functions. In
  this case, the solution for (2.6) or (2.7) is given by
\begin{equation}
 {T_1}=(S_{1}a+S_{2}b)R_{1}+(S_{1}d+S_{2}c)R_{3},
\end{equation}
\begin{equation}
{T_2}=(S_{1}a+S_{2}b)R_{2}+(S_{1}d+S_{2}c)R_{4},
\end{equation}
\begin{equation}
{T_3}=(S_{3}a+S_{4}b)R_{1}+(S_{3}d+S_{4}c)R_{3},
\end{equation}
\begin{equation}
{T_4}=(S_{3}a+S_{4}b)R_{2}+(S_{3}d+S_{4}c)R_{4},
\end{equation}
where a,b,c and d are constant $n \textbf{x}n$ matrices, and using
(2.21)-(2.24) in (2.4) we will have completed the mapping among
manifolds. In many situations where it is not possible to consider
$\frac{d{t}}{d\tau}, X_{lj}, Y_{lj}$ as explicit functions of one
of the two parameters, t or $\tau,$ we should expand them in
series of $\tau$, for example [9], so that with the modified
Hamiltonian formalism we can map one differential equation system
into another.
\renewcommand{\theequation}{\thesection.\arabic{equation}}
\section{\bf The Second Modified Formalism}
\setcounter{equation}{0}
 $         $

In the first modified Hamiltonian formalism only Hamilton
equations will be conserved, in the sense that they will be
transformed into other Hamilton equations by a non-canonical or
non-sympletic transformation, and the Poisson brackets will not be
invariant. The second modified formalism differs from the first
because we will have a set of signs in the Hamilton equations. We
use one class of functions  that include usual and unusual
Hamiltonians in both formalisms. We will maintain part of the
usual notation. Consider a time-dependent function $H({\tau})$
where ${\tau}$ is an affine parameter. Let us define 2n variables
that will be called ${\xi}^j$ with index j running from 1 to 2n so
that we have ${\xi}^j$ $\in$
$({\xi}^1,\ldots,{\xi}^n,{\xi}^{n+1},\ldots,{\xi}^{2n})$ =$(
{q}^1,\ldots,{q}^n,{p}^1,\ldots,{p}^n)$ where ${q}^j$ and ${p}^j$
can be or not the usual coordinates and momenta, respectively. We
now define the function by
\begin{equation}
 H({\tau})=\frac{1}{2}H_{ij}{\xi}^i{\xi}^j,
\end{equation}
where $H_{ij}$ is a  symmetric matrix. Consider the following
system
\begin{equation}
\frac{d{\xi}^i}{d\tau
}=I_{1}^{ik}\frac{\partial{H}}{\partial{\xi}^k } .
\end{equation}
The equation (3.2) introduces the $I_{1},$ given by
\begin{equation}
\left(%
\begin{array}{cc}
  O & A \\
  B & O \\
\end{array}%
\right)
\end{equation}
where O, A and B are the $n \textbf{x}n$, with O as the  zero
matrix, and  $A=\epsilon_{1}I$,   $B=\epsilon_{2}I$ are
proportional to identity matrix, with  $\epsilon_{i}=-1,+1$ and
$i=1$ or 2. We now make a linear transformation from ${\xi}^j$ to
${\eta}^j$ given by
\begin{equation}
  {\eta}^j={{T}^j}_k{\xi}^k,
\end{equation}
where ${{T}^j}_k$ is  a non-sympletic matrix, and the new function
is given by
\begin{equation}
 \bar{H}=\frac{1}{2}C_{ij}{\eta}^i{\eta}^j,
\end{equation}
where $C_{ij}$ is a  symmetric matrix. Consider that (3.5) obeys
the following equation
\begin{equation}
\frac{d{\eta}^i}{d\tau
}=I_{2}^{ik}\frac{\partial{H}}{\partial{\eta}^k },
\end{equation}
where $I_{2}$ is given by
\begin{equation}
\left(%
\begin{array}{cc}
  O & E \\
  D & O \\
\end{array}%
\right)
\end{equation}
and O, E and D are $n \textbf{x}n$, with O as the  zero matrix,
and  $E=\epsilon_{3}I$,   $D=\epsilon_{4}I$ are proportional to
the identity matrix, with  $\epsilon_{j}=-1,+1$ and $j=3$ or 4.
The functions A, B, E, and D could be chosen as arbitrary diagonal
matrices, however, such possibility will not be used in this
paper. The matrices H, C, and T obey the following system
\begin{equation}
 \frac{d{{T}^i}_j}{d\tau}+\frac{d{t}}{d\tau}{{T}^i}_kI_{1}^{kl}X_{lj}=I_{2}^{im}Y_{ml}{{T}^j}_k,
\end{equation}
where $2X_{lj}=\frac{\partial{H_{ij}}}{\partial{\xi}^l
}\xi^{i}+2H_{lj}$ and
$2Y_{ml}=\frac{\partial{C_{il}}}{\partial{\eta}^m
}\eta^{i}+2C_{ml},$ t and $\tau$ are the proper-times of the
particle in two different manifolds. We note that (3.8) is a first
order linear differential equation system in ${{T}^i}_k ,$ and
that the  non-linearity in the Hamiltonians  were transferred to
their coefficients. Consider $\frac{d{t}}{d\tau}X_{lj}=Z_{lj}$ and
write (3.8) in the matrix form
\begin{equation}
 \frac{d{T}}{d\tau}+TI_{1}Z=I_{2}YT,
\end{equation}
where T, Z and Y are  $2n \textbf{x}2n$ matrices as
\begin{equation}
\left(%
\begin{array}{cc}
  T_{1} & T_{2} \\
  T_{3} & T_{4} \\
\end{array}%
\right)
\end{equation}
with similar expressions for Z and Y. Let us  write (3.9) as
follows
\begin{equation}
 \dot{T_1}=\epsilon_{3}(Y_{3}T_{1}+Y_{4}T_{3})-\epsilon_{2}T_{2}Z_{1}-\epsilon_{1}T_{1}Z_{3},
\end{equation}
\begin{equation}
 \dot{T_2}=\epsilon_{3}(Y_{3}T_{2}+Y_{4}T_{4})-\epsilon_{2}T_{2}Z_{2}-\epsilon_{1}T_{1}Z_{4},
\end{equation}
\begin{equation}
 \dot{T_3}=\epsilon_{4}(Y_{1}T_{1}+Y_{2}T_{3})-\epsilon_{2}T_{4}Z_{1}-\epsilon_{1}T_{3}Z_{3},
\end{equation}
\begin{equation}
 \dot{T_4}=\epsilon_{4}(Y_{1}T_{2}+Y_{2}T_{4})-\epsilon_{2}T_{4}Z_{2}-\epsilon_{1}T_{3}Z_{4}.
\end{equation}
Now consider
\begin{equation}
 \dot{S_1}=\epsilon_{3}(Y_{3}S_{1}+Y_{4}S_{3}),
\end{equation}
\begin{equation}
 \dot{S_2}=\epsilon_{3}(Y_{3}S_{2}+Y_{4}S_{4}),
\end{equation}
\begin{equation}
 \dot{S_3}=\epsilon_{4}(Y_{1}S_{1}+Y_{2}S_{3}),
\end{equation}
\begin{equation}
 \dot{S_4}=\epsilon_{4}(Y_{1}S_{2}+Y_{2}S_{4}),
\end{equation}
and
\begin{equation}
 \dot{R_1}=-\epsilon_{2}R_{2}Z_{1}-\epsilon_{1}R_{1}Z_{3},
\end{equation}
\begin{equation}
 \dot{R_2}=-\epsilon_{2}R_{2}Z_{2}-\epsilon_{1}R_{1}Z_{4},
\end{equation}
\begin{equation}
 \dot{R_3}=-\epsilon_{2}R_{4}Z_{1}-\epsilon_{1}R_{3}Z_{3},
\end{equation}
\begin{equation}
 \dot{R_4}=-\epsilon_{2}R_{4}Z_{2}-\epsilon_{1}R_{3}Z_{4}.
\end{equation}
From the theory of  first order differential equation systems [9],
  it is well-known that each system in (3.15)-(3.22) has a solution in the
  region where $Z_{lj}$ and $Y_{ml}$ are continuous functions. In
  this case, the solution for (3.8) or (3.9) is given by
\begin{equation}
 {T_1}=(S_{1}a+S_{2}b)R_{1}+(S_{1}d+S_{2}c)R_{3},
\end{equation}
\begin{equation}
{T_2}=(S_{1}a+S_{2}b)R_{2}+(S_{1}d+S_{2}c)R_{4},
\end{equation}
\begin{equation}
{T_3}=(S_{3}a+S_{4}b)R_{1}+(S_{3}d+S_{4}c)R_{3},
\end{equation}
\begin{equation}
{T_4}=(S_{3}a+S_{4}b)R_{2}+(S_{3}d+S_{4}c)R_{4},
\end{equation}
where a,b,c and d are constant $n \textbf{x}n$ matrices, and using
(3.23)-(3.26) into (3.4) we will have completed the mapping among
manifolds. Using the first or the second formalism we can build
maps among manifolds. They are not equivalent  maps among
manifolds and the choice of one of them is not a preference
matter, but the second formalism can be reduced to the first one
by an appropriate choice of the constants $\epsilon_{i}$ and
$\epsilon_{j}$ . As in the first formalism, it is important to
note that the same particle has different proper-times in
different manifolds, so that line elements are not preserved by
local non-sympletic maps among manifolds. The derivative
$\frac{d{t}}{d\tau}$ increases the difficulty in (3.8), so that we
assume  the condition $\frac{d{t}}{d\tau}=1$. It implies  in a
decrease on mapped regions. The local non-sympletic maps are
well-defined for equal proper-times and time intervals. In this
paper, for the same particle in different manifolds with different
proper-times, we use the proper-time of one of the manifolds, so
that (3.8) assumes the following form
\begin{equation}
 \frac{d{{T}^i}_j}{d\tau}+{{T}^i}_kI_{1}^{kl}X_{lj}=I_{2}^{im}Y_{ml}{{T}^j}_k.
\end{equation}
As a consequence (3.23)-(3.26) will be simplified. It is important
to note that all that we call Hamiltonian, sometimes  are not true
Hamiltonians because they are not usual functions of coordinates
and momenta.
\renewcommand{\theequation}{\thesection.\arabic{equation}}
\section{\bf Spaces Associated to Spinors }
 \setcounter{equation}{0}
$        $

In this section we assume the convention used in [10] and we will
use the second modified formalism, although we could use the first
one. In this and in the following sections, what we call
Hamiltonian are not true Hamiltonians because they are not usual
functions of coordinates and momenta. In other words, we have
generic spaces. We note that in the first and  second formalisms
$\tau$ can be one parameter without association to a particle or
to any physics question. Let us consider
\begin{equation}
H({\tau})=\frac{1}{2}H_{ij}{\xi}^i{\xi}^j=\frac{1}{2}(\bar{X}^{t}
M^{t}X+X^{t}M\bar{X})
\end{equation}
where ${\xi}^j$ $\in$ $(
{\xi}^0,\ldots,{\xi}^n,{\xi}^{n+1},\ldots,{\xi}^{2n+2})$ =$(X^{0},
{X}^1,\ldots,{X}^n,\bar{X}^{1},\ldots,\bar{X}^{n+1}),$ and
$X^{0},$ ${X}^j$ and ${\bar{X}}^j$ are coordinates in a
2n+2-dimensional manifold. Sometimes they can be identified as
complex and complex-conjugated coordinates, respectively. We have
that $H_{ij}$ is a symmetric matrix given by
\begin{equation}
\left(%
\begin{array}{cc}
  O & M \\
  M^{t} & O \\
\end{array}%
\right)
\end{equation}
where O is the $(n+1) \textbf{x}(n+1)$ zero matrix, $M^{t}$ is the
$(n+1) \textbf{x}(n+1)$ transposed matrix of M. Explicitly
\begin{equation}
H({\tau})=\frac{1}{2}H_{ij}{\xi}^i{\xi}^j=\bar{X}^{i} M_{ij}X^{j},
\end{equation}
where $M_{ij}$ can be complex, having  or not a defined symmetry.
Using the Hamiltonian (4.1) in (3.2), we obtain
\begin{equation}
\frac{d{X}^i}{d\tau}=\epsilon_{1}\frac{\partial{H}}{\partial{\bar{X}}^i},
\end{equation}
and
\begin{equation}
\frac{d{\bar{X}}^i}{d\tau}=\epsilon_{2}\frac{\partial{H}}{\partial{X}^i}.
\end{equation}
 Let us consider
\begin{equation}
 F=\frac{1}{2}C_{ij}{\eta}^i{\eta}^j=\frac{1}{2}(\bar{x}^{t}
N^{t}x+x^{t}N\bar{x})
\end{equation}
where ${\eta}^j$ $\in$ $(
{\eta}^{0},\eta^{1},\ldots,{\eta}^{n},{\eta}^{n+1},\ldots,{\eta}^{2n+2})$
=$(
x^{0},x^{1},\ldots,x^{n},\bar{x}^{0},\bar{x}^{1},\ldots,\bar{x}^{n})$
and
 $x^{j}$ and $\bar{x}^{j}$ are coordinates in another 2n+2-dimensional manifold.
 We have that $C_{lk}$ is a symmetric matrix given by
\begin{equation}
\left(%
\begin{array}{cc}
  O & N \\
  N^{t} & O \\
\end{array}%
\right)
\end{equation}
where O is the $(n+1) \textbf{x}(n+1)$ zero matrix, $N^{t}$ is the
$(n+1) \textbf{x}(n+1)$ transposed matrix of N, and
$N_{lk}=\delta_{kl}.$ Explicitly
\begin{equation}
 F=\bar{x}^{0}x^{0}+\bar{x}^{1}x^{1}+ \ldots +\bar{x}^{n}x^{n},
\end{equation}
where (4.8) is a 2n+2-dimensional manifold. Using (3.23)-(3.26)
into (3.4) we will have a map between (4.1) and (4.6). For
$\bar{x}^{0}=x^{0}$ we have a 2n+1-dimensional manifold
\begin{equation}
 \tilde{F}=(x^{0})^{2}+{\bar{x}}^{1}x^{1}+ \ldots
 +{\bar{x}}^{n}x^{n},
\end{equation}
where associated  spinors can be built [10]. For
$\bar{x}^{0}=x^{0}=0,$ we have the 2n-dimensional manifold
\begin{equation}
\tilde{F}=\bar{x}^{1}x^{1}+ \ldots +\bar{x}^{n}x^{n}.
\end{equation}
As in (4.9), we can build spinors associated to (4.10). We note
that (4.8) and (4.10) are not usual Hamiltonians, they are the
special forms chosen by Cartan [10].
\renewcommand{\theequation}{\thesection.\arabic{equation}}
\section{\bf Kaehler Manifolds}
\setcounter{equation}{0} $         $

In this section we present some facts about Kaehler manifolds and
use  the second formalism  to build maps among manifolds.

Let us consider a real 2n-dimensional manifold. We will denote the
coordinates of a point P by
$(x^{1},\ldots,x^{n},\bar{x}^{1},\ldots,\bar{x}^{n}),$ and build a
n-dimensional complex manifold, where P has the following complex
and complex-conjugated coordinates
\begin{equation}
z^{\alpha}=x^{\alpha}+i\bar{x}^{\alpha},
\end{equation}
\begin{equation}
\bar{z}^{\alpha}=x^{\alpha}-i\bar{x}^{\alpha},
\end{equation}
where $\alpha$ $\in$ $(1,\ldots,n)$. Consider a symmetric tensor
$g_{ij}$. It is self-adjoint if it satisfies
\begin{equation}
 g_{\bar{\alpha}\bar{\sigma}}= g_{\bar{\sigma}\bar{\alpha}}=\bar{g_{\alpha\sigma}}=\bar{g_{\sigma\alpha}}
\end{equation}
\begin{equation}
 g_{\alpha\bar{\sigma}}=
 g_{\bar{\sigma}\alpha}=\bar{g_{\bar{\alpha}\sigma}}=\bar{g_{\sigma\bar{\alpha}}}.
\end{equation}
Moreover, if the inverse  contravariant tensor $g^{ij}$ defined by
$ g_{ik}g^{kj}=\delta_{i}^{j}$ exists, the usual Christoffel
symbols, the Riemann-Christoffel curvature tensor, the Ricci
tensor, and the scalar curvature are all self-adjoint [11]. We
assume now that, in this complex manifold, there is a positive
definite line element
\begin{equation}
 ds^{2}=g_{ij}dz^{i}dz^{j},
\end{equation}
where the symetric tensor $g_{ij}$ is self-adjoint and satisfies
\begin{equation}
 g_{\alpha\sigma}=g_{\bar{\alpha}\bar{\sigma}}=0,
\end{equation}
and (5.4). In this case the metric tensor is called a Hermitian
metric and the line element has the following expression
\begin{equation}
 ds^{2}=2g_{\alpha\bar{\sigma}}dz^{\alpha}d\bar{z}^{\sigma}.
\end{equation}
The Christoffel symbols are given by
\begin{equation}
 \Gamma^{\alpha}_{\mu\nu}=\frac{1}{2}g^{\alpha\bar{\sigma}}(\frac{\partial g_{\bar{\sigma}\mu}}{\partial
 z^{\nu}}+\frac{\partial g_{\bar{\sigma}\nu}}{\partial z^{\mu}}),
\end{equation}
\begin{equation}
 \Gamma^{\alpha}_{\mu\bar{\nu}}=\frac{1}{2}g^{\alpha\bar{\sigma}}(\frac{\partial g_{\mu\bar{\sigma}}}{\partial
 \bar{z}^{\nu}}-\frac{\partial g_{\mu\bar{\nu}}}{\partial \bar{z^{\sigma}}}),
\end{equation}
\begin{equation}
  \Gamma^{\alpha}_{\bar{\mu}\bar{\nu}}=0,
\end{equation}
and other components are given by symmetry and self-adjointness.
The components (5.9) transform as tensors. Kaehler  made the
following choice
\begin{equation}
 \Gamma^{\alpha}_{\mu\bar{\nu}}=0.
\end{equation}
It is the same as
\begin{equation}
\frac{\partial g_{\mu\bar{\sigma}}}{\partial
 \bar{z}^{\nu}}=\frac{\partial g_{\mu\bar{\nu}}}{\partial
 \bar{z^{\sigma}}},
\end{equation}
\begin{equation}
\frac{\partial g_{\bar{\mu}\sigma}}{\partial
  z^{\nu}}=\frac{\partial g_{\bar{\mu}\nu}}{\partial z^{\sigma}},
\end{equation}
so that
\begin{equation}
 g_{\bar{\sigma}\alpha}=\frac{\partial ^{2} \phi}{\partial z^{\alpha}\partial
 \bar{z}^{\sigma}}.
\end{equation}
The self-adjointness of $g_{\alpha\bar{\sigma}}$ implies that
$\phi$ is a real valued function. The (5.14) is called Kaehler
condition and a metric satisfying (5.4), (5.6) and (5.14) will be
called a Kaehler metric. Thus, in a Kaehler metric we have
\begin{equation}
 \Gamma^{\alpha}_{\mu\nu}=g^{\alpha\bar{\sigma}}\frac{\partial g_{\bar{\sigma}\mu}}{\partial
 z^{\nu}},
\end{equation}
\begin{equation}
 \Gamma^{\bar{\alpha}}_{\bar{\mu}\bar{\nu}}=g^{\bar{\alpha}\sigma}\frac{\partial g_{\sigma\bar{\mu}}}{\partial
  \bar{z}^{\nu}},
\end{equation}
and Riemann-Christoffel tensor components will be simplified. For
a n-dimensional Kaehler manifold, if at every point, the sectional
curvature is the same for all possible 2-dimensional sections,
then the curvature tensor is identically zero. The same is not
true for the holomorphic sectional curvature, thus, if we assume
that at all points of the manifold they are all the same, we have
\begin{equation}
 R_{\alpha\bar{\sigma}\mu\bar{\nu}}=\frac{K}{2}(g_{\alpha\bar{\sigma}}g_{\mu\bar{\nu}}
 +g_{\alpha\bar{\nu}}g_{\mu\bar{\sigma}}).
\end{equation}
From (5.17)
\begin{equation}
 R_{\alpha\bar{\sigma}}=R_{\bar{\sigma}\alpha}=\frac{(n+1)K}{2}g_{\alpha\bar{\sigma}},
\end{equation}
where (5.18) is an Einstein manifold.

 Let us consider
\begin{equation}
H({\tau})=\frac{1}{2}H_{ij}{\xi}^i{\xi}^j=\frac{1}{2}(\bar{x}^{t}
M^{t}x+x^{t}M\bar{x})
\end{equation}
where ${\xi}^j$ $\in$ $(
{\xi}^1,\ldots,{\xi}^n,{\xi}^{n+1},\ldots,{\xi}^{2n})$ =$(
{x}^1,\ldots,{x}^n,{\bar{x}}^1,\ldots,{\bar{x}}^n)$ and ${x}^j$
and ${\bar{x}}^j$ are real coordinates in a 2n-dimensional
manifold. We have that $H_{ij}$ is a symmetric matrix given by
\begin{equation}
\left(%
\begin{array}{cc}
  O & M \\
  M^{t} & O \\
\end{array}%
\right)
\end{equation}
where O is the $n \textbf{x}n$ zero matrix, $M^{t}$ is the $n
\textbf{x}n$ transposed matrix of M. Using the Hamiltonian (5.19)
in (3.2), we obtain
\begin{equation}
\frac{d{x}^i}{d\tau}=\epsilon_{1}\frac{\partial{H}}{\partial{\bar{x}}^i},
\end{equation}
and
\begin{equation}
\frac{d{\bar{x}}^i}{d\tau}=\epsilon_{2}\frac{\partial{H}}{\partial{x}^i
}.
\end{equation}
We can present a new Hamiltonian similar to (5.19) and build a map
between them, and consider that they are real representations of
two complex manifolds. We still have another possibility  where we
present two Hamiltonians  with complex and complex-conjugated
coordinates given by (5.1) and (5.2). For this last option we can
build a map between (5.7) and another Kaehler line element.
\section{Concluding Remarks}
  $              $
The objective of this paper is twofold. Firstly, it presents a
second modified formalism as an option  to and a generalization of
the first one [6]. Secondly, it shows, through maps, the use of
this second formalism in some areas of mathematics such spaces
associated to spinors and complex spaces. Explicit applications
will be presented in a next paper. We could have presented a
section with maps among Finsler spaces. For such, it would be
necessary the usual choice of a metric tensor with appropriate
momentum dependence. However a more general or not momentum
dependence can be introduced directly in (2.7) or (3.9). It would
be a repetitive procedure, therefore, we decided not to include
it. Invariance is a fundamental property in many theories, as in
general relativity. However, if we want to build maps among
manifolds, the first or the second modified formalism can be
useful, and both can be considered additional  mathematical
resources for research.

\end{document}